\documentclass{article}

\usepackage{PRIMEarxiv}
\usepackage{amsmath}
\usepackage[utf8]{inputenc} 
\usepackage[T1]{fontenc}    
\usepackage{hyperref}       
\usepackage{url}            
\usepackage{booktabs}       
\usepackage{amsfonts}       
\usepackage{nicefrac}       
\usepackage{microtype}      
\usepackage{lipsum}
\usepackage{fancyhdr}       
\usepackage{graphicx}       
\graphicspath{{media/}}     

\title{FuseCaps: Investigating Feature Fusion Based
Framework for Capsule Endoscopy Image Classification}

\author{
  Bidisha Chakraborty \\
  Government College of Engineering and Leather Technology \\
  Kolkata, India\\
  \texttt{bidisha23102000@gmail.com} \\
   \And
  Shree Mitra \\
  Indian Institute of Information Technology Guwahati \\
  Guwahati, India\\
  \texttt{shree.mitra23m@iiitg.ac.in} \\
}

\begin{document}
\maketitle

\begin{abstract}
In order to improve model accuracy, generalization, and class imbalance issues, this work offers a strong methodology for classifying endoscopic images. We suggest a hybrid feature extraction method that combines convolutional neural networks (CNNs), multi-layer perceptrons (MLPs), and radiomics. Rich, multi-scale feature extraction is made possible by this combination, which captures both deep and handmade representations. These features are then used by a classification head to classify diseases, producing a model with higher generalization and accuracy. In this framework we have achieved a validation accuracy of 76.2\% in the capsule endoscopy video frame classification task.
\end{abstract}


\section{Introduction}
Endoscopy is a procedure that is used to have a close look at the organs. This is done either to detect the diseases or observe the cellular patterns. Early detection of diseases is important as it helps to reduce the mortality rate and also improve the development of medicines. Here the subset of Machine Learning that is Deep Learning comes into play. Deep Learning has been implemented widely in the medical field to detect gastrointestinal and liver-related diseases. As a result, many models have been developed to classify capsule endoscopy images. Some of them involve the use of Convolution Neural Networks or Transfer Learning. But to ensure that the classification of the images is to the point, we have introduced the combination of Radiomics and Convolution Neural networks to enrich the feature dataset and lastly use that feature dataset to perform classification.
\section{Methods}

In general a wide variety of techniques are used to determine the category of diseases. Some of them are Supervised Learning, Transformation Learning, Convolutional Neural Networks and so on. The above defined techniques have been also combined to enrich the feature dataset. 

In this we have proposed a methodology that will not only increase the accuracy of the model but will also improve the generalisation of the model and will also handle class imbalance. For extraction of features from the dataset, we have implemented a combination of Radiomics followed by Multi Layer Perceptron and Convolutional Neural Networks. The classification head is used for classification of diseases.

\subsection{Radiomics Feature Extraction}
Radiomics\cite{LAMBIN2012441} is a process in which we try to extract quantitative or handcrafted features from the images. This method is particularly useful in the medical field as medical images has wide variety of handcrafted features including shape, texture, density etc and this technique largely helps to derive those features using data characterisation algorithms. The algorithms also makes use of advanced mathematical concepts like Laplacian, Gaussian, Gradient formulas to apply the filters. Radiomics also makes use of advanced statistical methods like Gray Level co-occurrence matrix, Gray Level Run Length matrix, Gray Level Zone Size Matrix etc. But we Radiomics is particularly useful in segmentation where the Region of Interest is masked and the foreground is present but unmasked.
\paragraph{}In our case since there was no foreground present, so first we considered the centre of image and converted the pixel values in the central region to 1 while the rest to 0. We applied the Radiomics to extract features from the central part of the image. The results are stored in NPY file and is ultimately concatenated in a Comma Separated file that has features like mean, version etc.
Now for the second case we are extracting features from the sides and excluding the central portion of the image. So basically we are creating our own foregrounds as per our needs, masking the Region of Interest and accordingly extracting the features.
\paragraph{}After concatenating the results, we combine the two csv files side by side and dropped the unnecessary columns. After that it is passed through the Multi Layer Perceptron.

\subsection{Multi Layer Perceptron}
Multi-Layer Perceptrons\cite{haykin1994neural} are Artificial Neural Networks that comprise an input layer, one or more hidden layers, and output layers. We used a multi-layer perceptron to preprocess the handcrafted features further. \par The MLP maps an input radiomics vector \( \mathbf{x} \in \mathbb{R}^{d_{\text{in}}} \) to a compact embedding \( \mathbf{z} \in \mathbb{R}^{d_{\text{embed}}} \) by first applying a linear transformation, 
\[
\mathbf{h}_1 = \text{ReLU\cite{DBLP:journals/corr/abs-1803-08375}}(\mathbf{W}_1 \mathbf{x} + \mathbf{b}_1),
\]
where \( \mathbf{W}_1 \in \mathbb{R}^{1024 \times d_{\text{in}}} \). Dropout\cite{10.5555/2627435.2670313} regularization, 
\[
\mathbf{h}_2 = \text{Dropout\cite{10.5555/2627435.2670313}}(\mathbf{h}_1, p=0.5),
\]
mitigates overfitting by randomizing feature selection. Another linear transformation, 
\[
\mathbf{z} = \text{Dropout\cite{10.5555/2627435.2670313}}(\mathbf{W}_2 \mathbf{h}_2 + \mathbf{b}_2, p=0.5),
\]
with \( \mathbf{W}_2 \in \mathbb{R}^{d_{\text{embed}} \times 1024} \), completes the embedding. This process reduces dimensionality and enhances the representation of complex, non-linear feature relationships, producing a robust embedding that can be effectively fused with CNN features for multi-modal classification of capsule endoscopy images.

\subsection{Convolutional Neural Networks}
For image classification tasks or processing tasks we mainly use Convolutional Neural Networks or CNNs\cite{LeCun2015}. The backbone CNN model of the proposed framework is the DenseNet\cite{Huang_2017_CVPR} CNN architecture, which we used to extract features from the dataset's complicated endoscopic pictures. DenseNet's ability to retrieve feature maps from all previous levels is the primary driving force behind its use. Dense blocks, in which the output of every layer inside a block is concatenated with all of the inputs of its preceding layers, are used to accomplish this. In particular,
\begin{equation}
X_l = H_l([X_0; X_1; \ldots ; X_{l-1}]) \in \mathbb{R}^{h_b \times w_b \times d_b}
\label{eq:xl}
\end{equation}

where \( X_0, X_1, \ldots, X_l \in \mathbb{R}^{h_b \times w_b \times d_b} \) represent the output feature maps from the \( 0 \)-th to the \( l \)-th layers, ";" denotes the concatenation operation, \( d_b \) is the feature dimension of the dense block \( b \), and the convolution function \( H_l(\cdot) \) consists of a 3 × 3 convolution layer, a ReLU activation function, and batch normalization.

\subsection{Projection Head}

In the context of image classification, let \( x \in \mathcal{X} \subset \mathbb{R}^{H \times W \times C} \) be the input image and \( f_{\theta}(x) = z \in \mathcal{F} \subset \mathbb{R}^{h \times w \times d} \) represent the high-dimensional feature vector obtained from a convolutional neural network (CNN). The dimensionality of \( z \) is often large, leading to potential redundancies that can impede the learning of discriminative features. To mitigate this, a projection head \( g_{\phi}(z) = y \in \mathbb{R}^{k} \) is employed, where \( k \ll h \times w \times d \).

The projection head\cite{10.5555/3524938.3525087} incorporates an Adaptive Average Pooling operation, which reduces spatial dimensions while preserving essential information:
\[
z' = \text{GAP}(z) \in \mathbb{R}^d.
\]
Following this, a fully connected layer with Batch Normalization and ReLU activation transforms the pooled representation:
\[
y = \sigma(\textbf{W} z' + b) \in \mathbb{R}^k,
\]
where \( \sigma \) denotes the activation function. This transformation enhances feature disentanglement and regularizes the representation space, which is crucial for effective generalization.

By constraining the representation to a lower-dimensional space, the projection head not only reduces the risk of overfitting but also helps the classifier focus on relevant features, minimizing the empirical loss:
\[
\mathcal{L} = \frac{1}{N} \sum_{i=1}^N \ell(\hat{y}_i, y_i).
\]
Thus, the introduction of a projection head serves to amplify the discriminative power of the model while ensuring a compact, interpretable representation suitable for classification tasks.

\subsection{Integration of CNN Extracted Features and Numerical Radiomics Features}

Let \( \mathbf{X}_{img} \) be the image input, and \( \mathbf{X}_{num} \) be the numerical features derived from radiomics. The model outputs are represented as follows:

1. DenseNet Feature Extraction:
   \[
   \mathbf{F}_{cnn} = \text{DenseNet}(\mathbf{X}_{img})
   \]

2. MLP Feature Extraction:
   \[
   \mathbf{F}_{mlp} = \text{MLP}(\mathbf{X}_{num})
   \]

3. Projection Head for Dimensionality Reduction:
   \[
   \mathbf{F}_{proj} = \text{ProjectionHead}(\mathbf{F}_{cnn})
   \]

4. Feature Concatenation:
   The combined feature vector is given by:
   \[
   \mathbf{F}_{combined} = \mathbf{F}_{proj} \oplus \mathbf{F}_{mlp}
   \]
   where \( \oplus \) denotes the concatenation operation.

5. Classification Output:
   The final classification output is obtained by passing the combined features through the classifier:
   \[
   \mathbf{y} = \text{ClassificationHead}(\mathbf{F}_{combined})
   \]

\par The combined features capture both visual (\( \mathbf{F}_{cnn} \)) and numerical (\( \mathbf{F}_{mlp} \)) information, enhancing the representation power:
   \[
   \mathcal{L} = \text{Loss}(\mathbf{y}, \mathbf{y}_{true})
   \]
By integrating features, the model can learn a more complex decision boundary \( f(\mathbf{F}_{combined}) \) that separates different classes more effectively.
The integration leads to improved generalization on unseen data, represented mathematically as:
   \[
   \mathbb{E}[\mathcal{L}_{test}] < \mathbb{E}[\mathcal{L}_{train}]
   \]


\begin{figure}[htbp]
    \centering
    \includegraphics[width=0.9\linewidth]{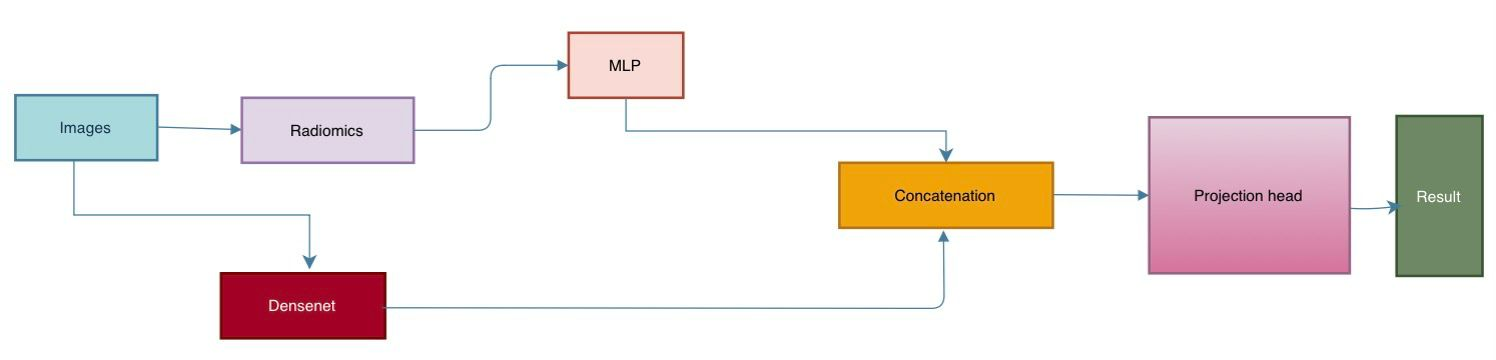}
    \caption{Block diagram of the developed pipeline.}
    \label{fig:enter-label}
\end{figure}

\section{Implementation details}
In our implementation, DenseNet is divided into three blocks, each of which has sixteen bottleneck levels in the encoder. When $\theta$ = $0.5$ is applied, this transition layer shrinks the channel and spatial sizes of the feature maps between each of the DenseNet's two blocks. The growth rate is $k = 24$, and the dropout rate is $p = 0.2$. 
We have employed an MLP\cite{haykin1994neural} with two linear layers of out-feature shapes 1024 and embedding size, respectively, as a non-linear projection head. This projection head provides an output feature of shape ($N$, embedding size), where $N$ is the batch size, and is combined with the densenet feature extractor. \par  We use Adam\cite{kingma2017adammethodstochasticoptimization} as our optimizer, with a learning rate of $1e-3$ and a weight decay of $1e-6$. The batch size is set at 64. 

\section{Results}
\begin{figure}[ht]
    \centering
    \includegraphics[width=0.5\linewidth]{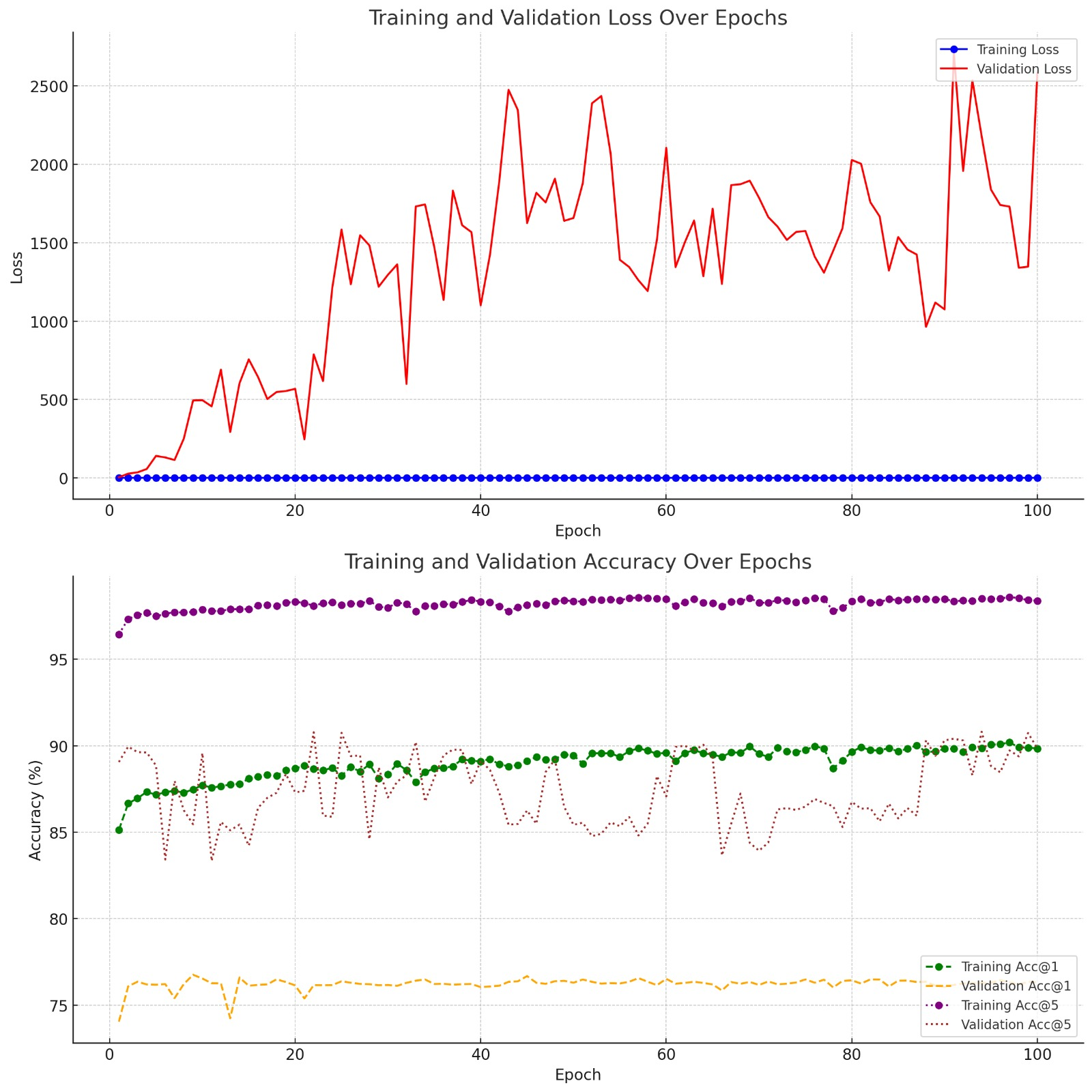}
    \caption{Training and Validation Loss and Accuracy Over Epochs}
    \label{fig:training_validation}
\end{figure}

The graphs in \textbf{Figure \ref{fig:training_validation}} depict the training and validation performance over 100 epochs for a model.

Overall, the results still indicates some chances of overfitting, where the model tries to generalize the images despite good training performance.

\subsection{Achieved results on the validation dataset}
\begin{figure}[ht]
    \centering
    \includegraphics[width=0.4\linewidth]{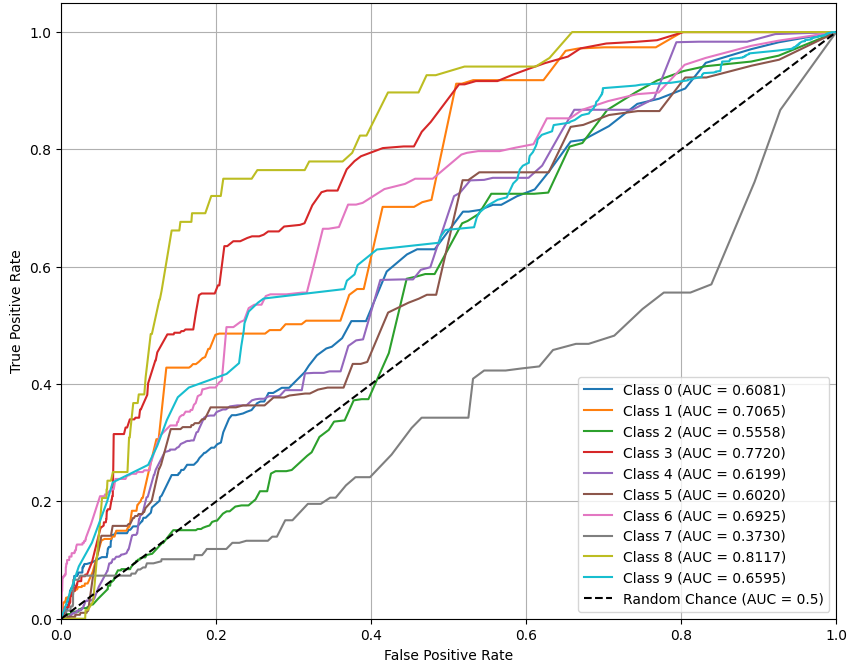}
    \caption{AUC-ROC Curve}
    \label{fig:roc}
\end{figure}

In \textbf{Figure \ref{fig:roc}}, the ROC curves and AUC scores for each class demonstrate how our model handles class imbalance. The AUC metric provides a threshold-independent assessment, which is less sensitive to class distribution than metrics like accuracy. Notably, despite class imbalance, certain minority classes (e.g., Class 3 and Class 8, with AUCs of 0.7720 and 0.8117, respectively) show high discriminatory power, indicating effective separation from other classes. Lower AUCs, such as for Class 7 (AUC = 0.3730), highlight areas for improvement. Overall, the per-class AUCs suggest that our approach maintains a balanced performance across both majority and minority classes, thereby partially mitigating the effect of class imbalance.




\begin{table*}[ht!]
\centering
\caption{Performance Comparison of Different Methods}
\label{tab:results}
\resizebox{\textwidth}{!}{
\begin{tabular}{lcccc}
\hline
\textbf{Method} & \textbf{Avg. ACC} & \textbf{Avg. Sensitivity} & \textbf{Avg. F1-score} & \textbf{Avg. Precision} \\ \hline
\textbf{Custom CNN (baseline)} & 0.460 & 0.097 & 0.093 & 0.100 \\
\textbf{ResNet50 (baseline)} & 0.760 & 0.320 & 0.373 & 0.602 \\
\textbf{SVM (baseline)} & 0.818 & 0.408 & 0.487 & 0.833 \\
\textbf{VGG16 (baseline)} & 0.568 & 0.543 & 0.484 & 0.525 \\ \hline
\textbf{Proposed Method} & \textbf{0.762} & \textbf{0.762} & \textbf{0.659} & \textbf{0.611} \\ \hline
\end{tabular}
}
\end{table*}

Here we have calculated the weighted average of the metrics.
As we can see our model performs well when compared with other models but we can expect our model to perform better which we can improve in the near future.
\section{Discussion}
As we can see the model is a bit over fitted due to huge class imbalance. In future we can improve the generalization of the model by introducing some more images of the minority classes. We can use GANs for synthetic generation of minority class images.

\section{Conclusion}
We have proposed a novel methodology in the field of capsule endoscopy video frame images for classification of diseases. We have achieved validation accuracy of 76.3\% but this can be further improved using more advanced techniques like Self-Supervised Learning, Synthetic Image Generation using GANs etc.
\section{Acknowledgments}
As participants in the Capsule Vision 2024 Challenge, we fully comply with the competition's rules as outlined in \cite{handa2024capsule}. Our AI model development is based exclusively on the datasets provided in the official release in \cite{Handa2024}.
\bibliographystyle{unsrt}  
\bibliography{references}

\begin{thebibliography}{10}

\bibitem{LAMBIN2012441}
Philippe Lambin, Emmanuel Rios-Velazquez, Ralph Leijenaar, Sara Carvalho, Ruud~G.P.M. {van Stiphout}, Patrick Granton, Catharina~M.L. Zegers, Robert Gillies, Ronald Boellard, André Dekker, and Hugo~J.W.L. Aerts.
\newblock Radiomics: Extracting more information from medical images using advanced feature analysis.
\newblock {\em European Journal of Cancer}, 48(4):441--446, 2012.

\bibitem{haykin1994neural}
Simon Haykin.
\newblock {\em Neural networks: a comprehensive foundation}.
\newblock Prentice Hall PTR, 1994.

\bibitem{DBLP:journals/corr/abs-1803-08375}
Abien~Fred Agarap.
\newblock Deep learning using rectified linear units (relu).
\newblock {\em CoRR}, abs/1803.08375, 2018.

\bibitem{10.5555/2627435.2670313}
Nitish Srivastava, Geoffrey Hinton, Alex Krizhevsky, Ilya Sutskever, and Ruslan Salakhutdinov.
\newblock Dropout: a simple way to prevent neural networks from overfitting.
\newblock {\em J. Mach. Learn. Res.}, 15(1):1929–1958, January 2014.

\bibitem{LeCun2015}
Yann LeCun, Yoshua Bengio, and Geoffrey Hinton.
\newblock Deep learning.
\newblock {\em Nature}, 521(7553):436--444, May 2015.

\bibitem{Huang_2017_CVPR}
Gao Huang, Zhuang Liu, Laurens van~der Maaten, and Kilian~Q. Weinberger.
\newblock Densely connected convolutional networks.
\newblock In {\em Proceedings of the IEEE Conference on Computer Vision and Pattern Recognition (CVPR)}, July 2017.

\bibitem{10.5555/3524938.3525087}
Ting Chen, Simon Kornblith, Mohammad Norouzi, and Geoffrey Hinton.
\newblock A simple framework for contrastive learning of visual representations.
\newblock In {\em Proceedings of the 37th International Conference on Machine Learning}, ICML'20. JMLR.org, 2020.

\bibitem{kingma2017adammethodstochasticoptimization}
Diederik~P. Kingma and Jimmy Ba.
\newblock Adam: A method for stochastic optimization, 2017.

\bibitem{handa2024capsule}
Palak Handa, Amirreza Mahbod, Florian Schwarzhans, Ramona Woitek, Nidhi Goel, Deepti Chhabra, Shreshtha Jha, Manas Dhir, Deepak Gunjan, Jagadeesh Kakarla, et~al.
\newblock Capsule vision 2024 challenge: Multi-class abnormality classification for video capsule endoscopy.
\newblock {\em arXiv preprint arXiv:2408.04940}, 2024.

\bibitem{Handa2024}
Palak Handa, Amirreza Mahbod, Florian Schwarzhans, Ramona Woitek, Nidhi Goel, Deepti Chhabra, Shreshtha Jha, Manas Dhir, Deepak Gunjan, Jagadeesh Kakarla, and Balasubramanian Raman.
\newblock {Training and Validation Dataset of Capsule Vision 2024 Challenge}.
\newblock {\em Fishare}, 7 2024.

\end{thebibliography}

\end{document}